\documentclass[12pt,preprintnumbers,aps,amssymb,nofootinbib,amsmath]{revtex4}
\usepackage{epsfig,epsf}
\usepackage{graphicx}
\usepackage{epstopdf}
\newcommand{\beq}{\begin{equation}}
\newcommand{\eeq}{\end{equation}}
\def\eq#1{{(\ref{#1})}}
\def\fig#1{{Fig.~\ref{#1}}}

\def\b#1{\mathbf{#1}}
\newcommand{\im}{\mathrm{Im}}
\newcommand{\ampl}{\mathcal{M}}

\newcommand{\e}{\varepsilon}
\begin{document} 

\preprint{ BNL-NT-06/43,\, RBRC-657}

\title{Vacuum self--focussing of very intense laser beams}

\author{Dmitri Kharzeev$^a$ and Kirill Tuchin$^{b,c}$}
 
\affiliation{
a) Department of Physics, Brookhaven National Laboratory,\\
Upton, New York 11973-5000, USA\\
b) Department of Physics and Astronomy,\\
Iowa State University, Iowa, 50011, USA\\
c) RIKEN BNL Research Center,\\
Upton, New York 11973-5000, USA\\
}

\date{\today}

\begin{abstract} 

We argue that long--range photon--photon 
attraction induced by the dipole interaction of two electron--positron loops can lead to 
``vacuum self-focussing" of very intense laser beams.  The focussing angle $\theta_F$ is found to 
increase with the beam intensity $I$ as $\theta_F \sim I^{4/3}$;  
for the laser beams available at present or in the near future, 
$\theta_F \simeq 10^{-10}\div10^{-7}$.

\end{abstract}

\maketitle

%%%%%%%%%%%%%%%%%%%%%%%%%%%%%%%%%%%%%%%%

\section{Introduction}

The response of quantum vacuum to strong external fields is a problem of great fundamental interest. 
Recent technological developments made available very intense laser beams, and opened the possibility 
to study non--linear QED interactions in experiment (for a recent  review, see  \cite{Marklund:2006my}). In particular, multi--photon effects have been already identified experimentally in light--by--light scattering \cite{Burke:1997ew}. 
Theoretical studies of light--by--light scattering so far have been limited to the one--fermion--loop effects -- 
the scattering of photons induced by the coupling to one electron--positron pair. The studies of higher loop corrections so far have been limited to two--loop effects arising from an additional internal virtual photon interacting with the single fermion loop; the corresponding results are available in an integral form \cite{Ritus}; see \cite{Dunne} for a comprehensive review\footnote{Some non-perturbative effects such as those caused by the existence of bound electron--positron states have also been considered \cite{Darewych:1992ar}.}.

The expansion in the number of fermion 
loops clearly converges very fast, since additional fermion loops bring into the effective low--energy interaction extra powers of $\alpha^2 F^2 / m^4 \ll 1$, where $F$ is the field strength tensor, $m$ is the electron mass and 
$\alpha \simeq 1/137$ is the electromagnetic coupling. 
We thus come to the conclusion that one can safely approximate 
the photon--photon scattering amplitude by a one-fermion-loop result. 

\begin{figure}[ht]
%\begin{center}
  \begin{tabular}{cc} 
      \includegraphics[width=5cm]{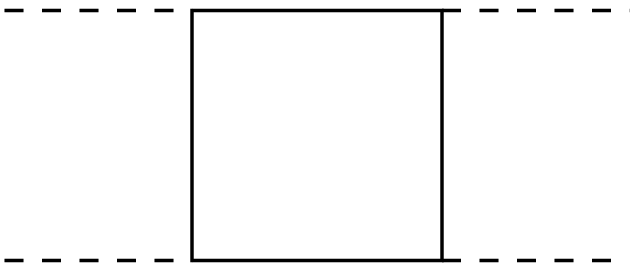} &
        \includegraphics[height=4cm]{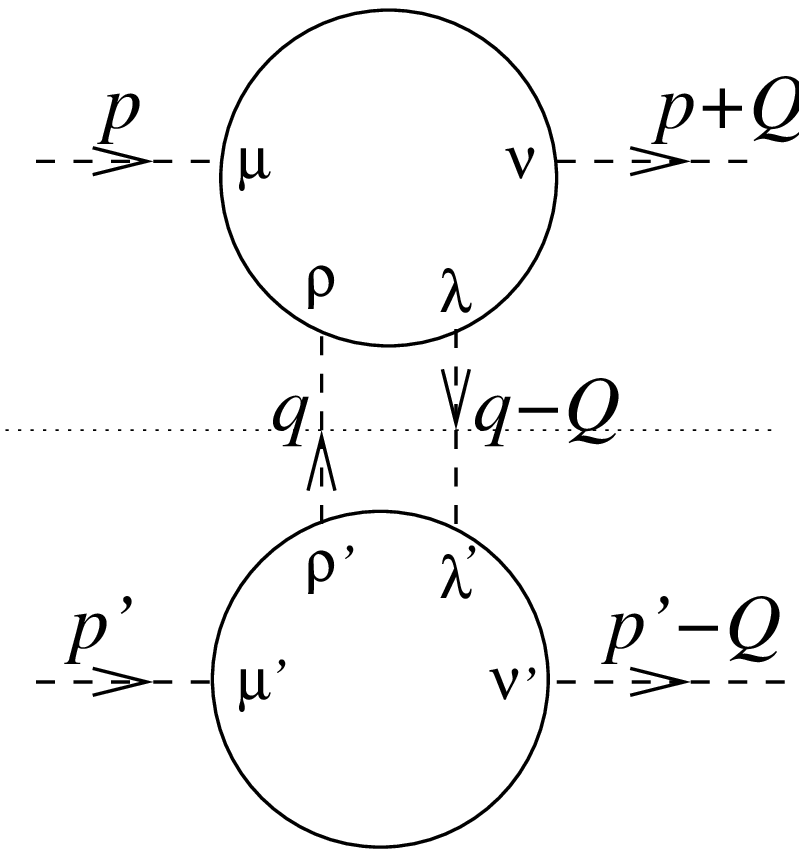}\\
         (a) & (b) 
\end{tabular}
%\end{center}
\caption{The amplitudes of light--by--light scattering at the one-fermion loop (a) and two-fermion-loop (b) level.}
\label{diag}
\end{figure}

This conclusion however may be premature for the field of an intense laser. 
Indeed, let us compare the photon--photon scattering amplitudes at the one--fermion--loop level (\fig{diag}(a)) and two--fermion--loop level (\fig{diag}(b)). The amplitude of the box diagram of \fig{diag}(a) 
in coordinate space falls off exponentially at large separations $r$ as $\sim (\alpha^2/m^4 r^7) \exp(- 2 m r )$ since it involves the exchange of a massive fermion pair in the $t-$channel. On the other hand, the amplitude 
of \fig{diag}(b) behaves in coordinate space like  $\sim \alpha^4/ m^{8}r^{11}$ (we will show this by an explicit calculation below).  The comparison of the amplitudes 
of \fig{diag}(a) and \fig{diag}(b) thus leads us to the conclusion that the two--fermion--loop amplitude of \fig{diag}(b) 
will dominate over the one--fermion--loop amplitude of \fig{diag}(a) at distances larger than 
\beq\label{cond}
r \sim -\ln\alpha/m.
\eeq 

One may wonder whether multi--photon interactions can also be important. There are two types of such interactions: {\it i)} when additional external beam photon lines are attached to a single fermion loop; 
and {\it ii)} when additional external beam photon lines  are attached 
to different fermion loops connected by $t-$channel photon pair  exchanges. The diagrams of type {\it i)} for $n$ external beam photon lines are suppressed by $(e F/ m^2)^n$ and fall off exponentially as a function of inter--photon separation; we thus expect that such interactions can be neglected when the field strength is smaller than the critical one, $F_c^2 \sim m^4/\alpha$.  The diagrams of type {\it ii)} are characterized by a power fall-off and so 
are potentially relevant even at sub--critical field strength; 
however the interactions involving $l$ beam photons are suppressed by $(\alpha^3/m^4 r^4)^l$. This parameter becomes of the order of unity at distances smaller than $r \sim \alpha^{3/4}/m$. In this paper we consider the dynamics of laser beams of smaller intensities, corresponding to larger inter--photon distances (note that $- \ln \alpha \gg \alpha^{3/4}$ for $\alpha = 1/137$),  see Eq.\eq{cond}.

This situation is completely analogous to the interaction of atoms: at short distances, the interaction 
involving the exchange of constituent electrons dominates, but at the distances large compared to the size of the 
atoms, the interaction is dominated by the (attractive) two--photon exchange 
despite the fact that it is of higher order in 
the coupling $\alpha$. The key question now is whether the conditions for the dominance of two--fermion--loop diagrams of \fig{diag}(b) can be realized in realistic laser beams.

Consider a laser beam which is characterized by the wavelength $\lambda$, coherence length $l_c$ and the transverse diameter $d$. The photon density in such a beam is determined by the ratio of the  field energy density to the energy of a single photon
\beq\label{est}
\rho\,=\,\frac{E^2}{\hbar\omega}\,,
\eeq
Let us approximate the beam by a cylinder of height $l$ and cross sectional area $S$ with the symmetry axis parallel to the beam direction. The beam energy contained in the cylinder is then $E^2 S l$. The beam intensity is the flux of energy through the unit surface, i.e. 
\beq
I\,=\,\frac{E^2 S l_c}{S(l_c/c)}=E^2 c\,,
\eeq 
hence
\beq\label{density}
\rho=\frac{1}{\hbar c}\frac{I}{\omega}\,.
\eeq
For example, for the Omega-EP laser at the Laboratory for Laser Energetics of the University of Rochester  the parameters are \cite{private}, $I=3\cdot 10^{20}$~W/cm$^2$ at the wavelength $\lambda=1054$~nm (corresponding to $\omega=2\pi c/\lambda=1.8\cdot 10^{15}$~s$^{-1}$). This amounts to the mean photon density of $\rho=5.3\cdot 10^{28}$~cm$^{-3}$. Therefore the typical inter-photon distance in the beam is  $ R\sim \rho^{-1/3}$ which is  $R=2.6\cdot 10^{-3}$~nm for the Omega-EP laser.

As long as the photons obey the linear Maxwell equations the laser beam is a coherent state. However, the coherent state can be destroyed by the mutual interaction of the photons. Such interactions appear as a result of quantum fluctuations of a photon into a virtual $e^+e^-$ pair. These pairs can be considered as electric or magnetic dipoles of size $\lambdabar_e=\hbar/m_ec=3.8\cdot 10^{-4}$~nm  which can  interact with virtual dipoles created by other photons in the beam. Thus, there are four dimensionful scales associated with the problem of photon interaction in the laser beam.  These are 
\beq\label{sep}
\lambdabar_e\,\ll\,  R\,\ll \lambda\,\ll L\,,
\eeq
where $L$ is the macroscopic length of the beam ($l_c$ or $d$). 
By comparing \eq{sep} and \eq{cond} we conclude that 
({\it i}) the condition of \eq{cond} for the dominance of two--fermion--loop interactions can be met in the available 
laser beams, and ({\it ii}) since the photon wavelength $\lambda$ is  the largest (microscopic) scale in the problem we can treat the two-photon interaction as a low--energy scattering problem.

%%%%
%%%%%

\section{Photon-photon interaction at large distances}

Let us now turn to the calculation of the photon--photon  potential density which can be expressed through the scattering amplitude $\mathcal{M}(s,Q^2)$ as
\beq\label{pot.fur}
V(\b r, t)=-\int \frac{d^4Q}{(2\pi)^4} \, e^{-i Q\cdot  x} \, \lim_{s\rightarrow 0}\ampl(s, Q^2)\,,
\eeq
where $x=(t,\b r)$ is the relative four-coordinate. This formula gives the potential density as a function of time. If we were considering the scattering of massive particles, we could have chosen their center-of-mass rest frame in which the potential would not depend on time $t$. This is not possible in the case of photon--photon interaction. Instead, we can obtain a meaningful quantity by averaging the potential \eq{pot.fur} over the period of the electromagnetic wave oscillation $T=2\pi/\omega$. We thus define the mean potential density as
\beq\label{aver}
\bar V(\b r)=\frac{1}{2 T}\int_{-T}^T\, V(\b r,t)\, dt\,.
\eeq

To calculate the two-photon interaction amplitude, \fig{diag}(b),  we use the $t$-channel dispersion relation 
\beq\label{disp}
\ampl(s,Q^2)\,=\, \frac{1}{\pi}\int \frac{d M^2}{M^2-Q^2-i0}\,\im \ampl(s,M^2)\,,
\eeq
where $M$ is the invariant mass of two $t$-channel photons. The imaginary part of the amplitude can be calculated by making the cut shown by the dotted line in \fig{diag}(b). This amounts to placing the cut photons on mass shell. According to the Mandelstam-Cutkosky rule the  cut photon's propagator is replaced by  
\beq\label{prop}
 \frac{-i g_{\mu\mu'}}{q^2+i0}\,\rightarrow 2\pi  \delta(p^2)\,
\sum _\mathrm{polariz.}\epsilon^*_\mu\epsilon_{\mu'}\,.
\eeq
Therefore, in the low energy limit 
\beq\label{imp}
\im \ampl(0,Q^2)\,=\, \int\frac{d^4q}{(2\pi)^2}
\delta(q^2)\delta((q-Q)^2)\, \sum _{\sigma\sigma'} \big| \ampl^\mathrm{box}_{++\sigma\sigma'}(M^2)\big|^2\,,
\eeq
where we used the fact that all photons in the beam have the same polarization (assigned $+$ here).

The low energy on--shell amplitudes corresponding to the box diagram \fig{diag}(a) can be found elsewhere (see e.g. in \cite{Berestetsky:1982aq}):
\beq\label{box}
\sum _{\sigma\sigma'} \big| \ampl^\mathrm{box}_{++\sigma\sigma'}(M^2)\big|^2\,=\, \bigg( \frac{11\alpha^2 M^4}{45 m^4}\bigg)^2\,+\, \bigg(- \frac{2\alpha^2 M^4}{15 m^4}\bigg)^2\,=\, 157\bigg( \frac{\alpha^2 M^4}{45 m^4}\bigg)^2\,.
\eeq
Taking the phase space integral in \eq{imp} and substituting it together with  \eq{box} into \eq{disp} yields the amplitude in the $t$-channel. However, we are interested in the $s$-channel amplitude. It can be obtained by analytic continuation to the region of space-like momentum $Q$. Therefore, the $s$-channel amplitude is 
\beq\label{m2}
\ampl(0, Q^2)\,=\, \frac{157}{8\pi^2} \, \bigg( \frac{\alpha^2}{45 m^4}\bigg)^2\, 
\int \frac{dM^2}{M^2-Q^2}\, M^8\,.
\eeq
where now $Q^2<0$. 
The potential density can be calculated by substituting \eq{m2} into  \eq{pot.fur}. We have
\beq\label{pot.fur1}
V(x)\,=\,-\frac{1}{(2\pi )^4}\frac{157}{8\pi^2} \, \bigg( \frac{\alpha^2}{45 m^4}\bigg)^2\, 
\int_0^\infty  dM^2 \,M^8\,\int d^4 Q \, e^{-i Q\cdot x}\, \frac{1}{M^2- Q^2-i0}\, .
\eeq
Integration over $Q$ yields
\beq\label{integ}
\int d^4 Q \, e^{i Q \cdot  x}\, \frac{1}{M^2- Q^2-i0}\,=\, 4\pi^2i \, \frac{M^2}{\sqrt{(-x^2+i0)M^2}}\, K_1(\sqrt{(-x^2+i0)M^2})\,,
\eeq
where $K_1$ is the modified Bessel function. Now we have to average the potential density \eq{pot.fur1} over the oscillation period according to \eq{aver}.  Since the field oscillation  period $T=\lambda$ is much larger than the inter-photon distance $R$ (the typical exchanged mass is  $M \sim R^{-1}$), we can set the integration limits in \eq{aver} to infinity.  The integral over $t$ can be done by rotating the integration contour by $\pi$ in the plane of complex $t$. The result is
\beq\label{potM}
\bar V(\b r)=
-\frac{1}{(2\pi )^4}\frac{157}{8\pi^2} \, \bigg( \frac{\alpha^2}{45 m^4}\bigg)^2\, 
\int_0^\infty  dM^2 \,M^8\, \frac{1}{2 T}\frac{4\pi^2}{r}\,e^{-Mr}\,.
\eeq
After integration over $M$  we finally derive
\beq\label{pot.com}
\bar V({\bf r})=-\frac{\omega}{(2\pi)^4}\frac{157}{16\pi^2}
\,\bigg( \frac{\alpha^2}{45 m^4}\bigg)^2\,
\frac{2\pi^2}{r^{11}}\,2\cdot 9!
\eeq
This is the mean potential density between two photons at large distances.

Let us compare this to the corresponding result for the interaction between atoms, where the potential falls off 
as $\sim r^{-6}$ at distances larger than the size of the atom $a$ but smaller than $a/\alpha$ (``Van der Waals interaction") and as $\sim r^{-7}$ at distances larger than  $a/\alpha$ due to relativistic retardation effects 
(``Casimir--Polder interaction")\footnote{The dispersive method which we use is analogous to the one used previously to describe the interaction of atoms and dipoles in Refs\cite{FS,Fujii:1999xn}.}. Our result falls off even faster, as $r^{-11}$, which can be traced back 
through the calculation to the gauge invariance of QED and the fact that the dipoles in our case are virtual, unlike 
in the interaction of atoms.    

%%%%
\section{Dispersion relation of the beam and self--focussing}

In this section we would like to calculate the refraction index $n$ of the intense laser beam taking into account the photon-photon attraction as described in the previous section. 
Assuming the dominance of two--particle interactions, the Hamiltonian describing the dynamics of interacting particles can 
be represented in the secondary quantization approach in the following general form (see e.g. \cite{BS}):
\begin{eqnarray}
\hat H&=& \sum \omega_{\b p}\, \hat a^\dagger_{\b p}\hat a_{\b p}+
\frac{1}{2}\sum \langle \b p,\b p'|\hat U|\b p+\b Q,\b p'-\b Q\rangle\, \hat a_{\b p}^\dagger \hat  a_{\b p'}^\dagger \hat a_{\b p+\b Q}\hat a_{\b p'-\b Q}\nonumber\\
&\approx & \sum \omega_{\b p}\, \hat a^\dagger_{\b p}\hat a_{\b p}+\frac{1}{2}\, U\, \sum_{\b p}
\hat a_{\b p}^\dagger \hat  a_{\b p}^\dagger \hat a_{\b p}\hat a_{\b p}\,.
\label{Hamil}
\end{eqnarray}
The  mean effective potential $U$ for scattering of two photons corresponding to the density \eq{pot.com} is given by: 
\begin{eqnarray}\label{Up}
 U&=&\int d^3 r\, \bar V(r)=\frac{1}{2T}\int_{-\infty}^\infty dt \int d^3 r\,\int \frac{d^4Q}{(2\pi)^4}\,e^{-iQ\cdot x}\,\ampl(0,Q^2)\nonumber\\
&=&- \frac{\omega}{4\pi}\,  \frac{157}{8\pi^2} \, \bigg( \frac{\alpha^2}{45 m^4}\bigg)^2\, 
\int_0^{1/R^2} dM^2\, M^8\,=\, -\frac{\omega}{4\pi} \frac{157}{8\pi^2} \, \bigg( \frac{\alpha^2}{45 m^4}\bigg)^2\, \frac{1}{4R^8}\,.\label{b}
\end{eqnarray}
The same result can be obtained directly by taking the four--dimensional integral of the right-hand-side of  \eq{integ}. Denoting by $4\pi A$ the forward elastic scattering amplitude we obtain from \eq{Up} $U=-A\, \omega$.

By analogy with the treatment of superfluidity we consider the quasi-particle fluctuations $\hat a_{\b p}=\hat c_{\b p}+\hat a'_{\b p}$ and ${\hat a}^\dagger_{\b p}=\hat c_{\b p}^\dagger+\hat {a'}_{\b p}^\dagger$ where the background  classical laser field is described by the commuting $c-$number operators $\hat c_{\b p}$ and $\hat c^\dagger_{\b p}$.  Expanding the Hamiltonian up to quadratic terms in  
$\hat a_{\b p}'$'s and diagonalizing the Hamiltonian yields 
\beq
\hat H\,=\, \omega N \,-\, \frac{1}{2}\,N_0\, N\,\omega\, A\,+\, \sum_{p}\e(\omega)\,\hat b_{\b p}^\dagger\hat b_{\b p}\,,
\eeq
where $\hat b_{\b p}$'s are the quasi-particle operators and $N_0$ is a mean number of particles in a sphere of radius $R$: $N_0\simeq N \cdot R^3/V\simeq \rho R^3 \simeq 1$. The dispersion law is given by
\beq\label{dispersion}
\e(\omega)\,=\, \omega\sqrt{1-A\,N_0/2}\simeq \omega\sqrt{1-A/2}\,.
\eeq

The redshift of the laser beam occurs since a fraction of the beam's kinetic energy density $E^2=\rho\, \omega$ (see \eq{est})   gets transformed into the interaction energy density  $(NN_0/2)\cdot U/V=U\, \rho/2$, where $N$ is the total number of photons  in the beam. Clearly, this fraction is $|U|/2\omega=A/2$.

The phase velocity  of the laser beam reads
\beq
u\,=\,\frac{\e(\omega)}{p} \approx 1-\frac{A}{4}\,.
\eeq
Eq.~\eq{b} implies that $A\sim R^{-8}\sim \rho^{8/3}$. 
The density of the beam $\rho$ decreases toward the beam edges. Therefore the beam phase velocity  $u$ increases towards the beam periphery.   This implies that the photon beam will self--focus: the refraction index $n=u^{-1}$ is largest in the center of the beam which has the ``focussing lens" effect. The focussing angle can be determined from the following simple observation. Focussing means that the photons in the beam center having phase velocity $1-A/4$ will arrive to the focussing point simultaneously with the photons at the beam periphery having unit phase velocity. This can only happen if the relative (``focussing") angle between these bunches of photons satisfies $\cos\theta_F=1-A/4$. Since the focussing angle is small we have
\beq\label{thetaF}
\theta_F\,=\,\sqrt{A/2}\,=\, \left( \frac{157}{16 \pi^3}\right)^{1/2} \, \bigg( \frac{\alpha^2}{180 m^4R^4}\bigg)\, .
\eeq

For the  Omega-EP laser beam we have the following estimate:  $\theta_F\simeq 1\cdot 10^{-10}$. We expect an increase in the focussing angle with the increase of the beam intensity according to the law $\theta_F\propto R^{-4}\propto \rho^{4/3}\propto I^{4/3}$, see \eq{b} and \eq{density}. The maximal possible effect can be achieved at $R\sim \lambdabar_e$ in which case $\theta_F \simeq 1\cdot 10^{-7}$. This value should be considered as an upper bound on the self--focussing angle due to the mechanism considered in this paper since our approach breaks down for $R \lesssim \lambdabar_e$. 
We have to emphasize that these estimates are admittedly qualitative, since  
they are derived in the mean field  approximation in which $ R\sim \rho^{-1/3}$.   

Our calculation shows that the focussing effect is present even for the plane wave, 
for which the leading order contribution from the box diagram of Fig.~1(a) vanishes.  
The magnitude of the expected effect however makes its experimental observation challenging. In particular, in a realistic experimental setup it would have to be distinguished from the non--linear effects caused by the presence of the air.  

%%%%%%%%%%%%%%%%%%%%%%%%%%%%%%%%%%%%%%%%

\vskip0.3cm
{\bf Acknowledgments}
%\vskip0.3cm
We are indebted to Adrian Melissinos and Yannis Semertzidis for the most useful and stimulating discussions. The work of D.K. was supported by the U.S. Department of Energy under Contract No. DE-AC02-98CH10886. 
K.T. would like to thank RIKEN, BNL and the U.S. Department of Energy (Contract
No. DE-AC02-98CH10886) for providing the facilities essential for the
completion of this work.

%%%%%%%%%%%%%%%%%%%%%%%%%%%%%%


\begin{thebibliography}{99}

%\cite{Marklund:2006my}
\bibitem{Marklund:2006my}
  M.~Marklund and P.~K.~Shukla,
  % ``Nonlinear collective effects in photon photon and photon plasma
  %interactions,''
  Rev.\ Mod.\ Phys.\  {\bf 78}, 591 (2006)
  [arXiv:hep-ph/0602123].
  %%CITATION = HEP-PH 0602123;%%


%\cite{Burke:1997ew}
\bibitem{Burke:1997ew}
  D.~L.~Burke {\it et al.},
  %``Positron production in multiphoton light-by-light scattering,''
  Phys.\ Rev.\ Lett.\  {\bf 79}, 1626 (1997).
  %%CITATION = PRLTA,79,1626;%%


\bibitem{Ritus}
V. I. Ritus, Zh. Eksp. Teor. Fiz. {\bf 69}, 1517 (1975) [ Sov. Phys. JETP {\bf 42}, 774 (1975) ]. 

%\cite{Dunne:2004nc}
\bibitem{Dunne}
  G.~V.~Dunne,
  %``Heisenberg-Euler effective Lagrangians: Basics and extensions,''
  arXiv:hep-th/0406216.
  %%CITATION = HEP-TH 0406216;%%

%\cite{Darewych:1992ar}
\bibitem{Darewych:1992ar}
  J.~W.~Darewych, M.~Horbatsch and R.~Koniuk,
  %``Photon Photon Resonances In Quantum Electrodynamics,''
  Phys.\ Rev.\ D {\bf 45}, 675 (1992).
  %%CITATION = PHRVA,D45,675;%%  
  

\bibitem{private} A. Melissinos, private communication. 

%\cite{Berestetsky:1982aq}
\bibitem{Berestetsky:1982aq}
  V.~B.~Berestetsky, E.~M.~Lifshitz and L.~P.~Pitaevsky,
 ``Quantum Electrodynamics''.
%\href{http://www.slac.stanford.edu/spires/find/hep/www?irn=1094556}{SPIRES entry}

\bibitem{FS}
G.~Feinberg and J.~Sucher, Phys.\ Rev. {\bf 139}, B 1619 (1965).

%\cite{Fujii:1999xn}
\bibitem{Fujii:1999xn}
  H.~Fujii and D.~Kharzeev,
  %``Long-range forces of {QCD},''
  Phys.\ Rev.\ D {\bf 60}, 114039 (1999)
  [arXiv:hep-ph/9903495].
  %%CITATION = HEP-PH 9903495;%%
  
\bibitem{BS}
N.N.~Bogoliubov and D.V.~Shirkov,  "Introduction to the theory of quantized fields", Interscience Publishers, 1949; Intersci.\ Monogr.\ Phys.\ Astron.\  {\bf 3}, 1 (1959).
 

\end{thebibliography}
\end{document}